# Effects of Cation Vacancy Distribution in Doped LaMnO$_{3+\delta}$ Perovskites

*Lorenzo Malavasi[a*], Clemens Ritter[b], Maria Cristina Mozzati[c], Cristina Tealdi[a], M. Saiful Islam[d], Carlo Bruno Azzoni[c], and Giorgio Flor[a]*

[a]Dipartimento di Chimica Fisica "M. Rolla", INSTM, IENI/CNR Unità di Pavia of Università di Pavia, V.le Taramelli 16, I-27100, Pavia, Italy.

[b]Institute Laue-Langevin, Boite Postale 156, F-38042, Grenoble, France.

[c]INFM, Unità di Pavia and Dipartimento di Fisica "A. Volta", Università di Pavia, Via Bassi 6, I-27100, Pavia, Italy.

[d]Chemistry Division, University of Surrey, Guildford, GU2 7XH, United Kingdom.

In this paper we report studies on the correlation between the presence and distribution of cation vacancies in doped manganites (La,M)MnO$_{3+\delta}$ (where M = Na, Ca) and their magnetic properties. Results indicate that cation vacancies are distributed differently for the different crystal structures and dopant ion type. In particular it is shown that knowledge of the total vacancies concentration alone is not enough to fully characterize the physical properties of manganites and that their distribution between the A and B sites of the perovskite structure plays a crucial role which should be taken into account in future studies.

*Corresponding Author: Dr. Lorenzo Malavasi, Dipartimento di Chimica Fisica "M. Rolla", INSTM, IENI/CNR Unità di Pavia of Università di Pavia, V.le Taramelli 16, I-27100, Pavia, Italy. Tel: +39-(0)382-987921 - Fax: +39-(0)382-987575 - E-mail: lorenzo.malavasi@unipv.it

# Introduction

In recent years there has been considerable interest in aliovalent doped LaMnO$_3$ perovskites (La$_{1-x}$M$_x$MnO$_3$ where M = Ca, Sr, Na) due to their negative magnetoresistance (MR) behaviour and the close correlation between structural, charge and orbital degrees of freedom [1-10]. The huge reduction in resistance found in these compounds when a magnetic field is applied is an example of Colossal Magnetoresistance (CMR). As revealed by many studies [11-18], the basic properties of mixed-valence manganites depend mainly on the relative amount of Mn(III) and Mn(IV) ions. The undoped LaMnO$_3$ compound, containing only Mn(III) ions, is in fact an insulating material with no MR effect and with an antiferromagnetic transition around 140 K. The reason for aliovalent doping is that it promotes oxidation of the Mn-array, which causes the material to became metallic and ferromagnetic at a defined temperature (the Curie temperature, $T_C$), which depends on the average Mn oxidation state. The optimal Mn(IV) concentration at which the highest $T_C$'s are found ranges from 25 to 30% [1,2].

More recently, it has also been pointed out that the point defects which depend on the oxygen content play a fundamental role in determining the physical properties of manganites [11,12,14-21]. When the usual synthetic routes are employed, the final products, particularly for $x < 0.3$, have an oxygen content different from the nominal value of 3 in La$_{1-x}$M$_x$MnO$_3$. Therefore, it appears that this variable has to be taken into serious consideration when studying the properties of the manganites.

When oxygen is in "excess" with respect to the ideal formula it is accommodated as structural oxygen and not on interstitial sites [23,24,26]; this can be expressed, according to Kröger-Vink notation for defect equilibria, as:

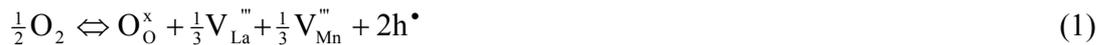

$$\tfrac{1}{2}O_2 \Leftrightarrow O_O^x + \tfrac{1}{3}V_{La}''' + \tfrac{1}{3}V_{Mn}''' + 2h^\bullet \qquad (1)$$

from which it follows that the oxygen excess or over-stoichiometry is compensated by cation vacancies ($V_{La}'''$, $V_{Mn}'''$) of equal concentration on the A and B sites of the perovskite structure. Atomistic simulation studies have also confirmed that Frenkel (interstitial) dopants are energetically unfavourable in the perovskite lattice [25].

The first investigation of cation vacancy distribution in oxygen over-stoichiometric manganites was carried by van Roosmalen *et al* [22]. Using transmission electron microscopy (TEM) this work suggested equal amounts of $V_{La}'''$ and $V_{Mn}'''$. Surprisingly, since then only a few studies, mostly involving neutron diffraction data of manganites, have considered this factor [20,21,27,28]. However it is clear that none of these studies have made a connection between the physical properties measured and the cation vacancy distribution. This is mainly due to the fact that the experiments were performed on only one composition so that no correlation with vacancy distribution could be easily extracted.

From the above discussion it is clear that the actual distribution of the cation vacancies between the A and B sites must strongly affect the physical properties of manganites. For example, cation vacancies on the manganese site should lead to a more pronounced deterioration of both the electrical and magnetic properties compared to vacancies on the lanthanum site, due to the blocking of the magnetic interaction paths and their influence on the Mn-O conduction band [11-13].

In our previous studies [11,12] on $(La_{1-x}Ca_x)_{1-\varepsilon}Mn_{1-\varepsilon}O_3$ (LCMO) and $(La_{1-x}Na_x)_{1-\varepsilon}Mn_{1-\varepsilon}O_3$ (LNMO) systems (where $\varepsilon = \delta/(3+\delta)$ with respect to the more familiar notation $LaMnO_{3+\delta}$) we proposed, in order explain the experimental results, that the cation vacancies in the two compositional series had different distributions. Our evidence, however, provided only an *indirect* correlation between the point defect concentration and distribution and the structural, magnetic and electrical data for the LCMO and LNMO systems. In the present work we try to obtain a *direct* correlation between the physical properties and the defect structures of these materials. Our aim is to stress that the actual cation vacancy concentration on the A and B sites of the perovskite structure must be considered (together with the tolerance factor, the average Mn oxidation state and the oxygen content) as a fundamental parameter in understanding the properties of perovskite manganites.

In order to provide a direct correlation we studied oxygen-overstoichiometric samples by means of neutron diffraction complemented by magnetic characterization. This was performed on three samples with nominally equal Mn(IV) contents: $La_{0.961\pm0.02}Mn_{0.961\pm0.02}O_3$, $(La_{0.95}Na_{0.05})_{0.974\pm0.02}Mn_{0.974\pm0.02}O_3$ and $(La_{0.9}Ca_{0.1})_{0.974\pm0.02}Mn_{0.974\pm0.02}O_3$ (here written in terms of equal numbers of vacancies on the A and B sites). These compounds were selected in an attempt to link the differences in the ratio of cation vacancies to differences in the crystal structure (the pure and the Na-doped samples are rhombohedral while the Ca-doped material is orthorhombic at room temperature, RT) and to the nature of the dopant ion. Finally, all the samples considered in this work have been prepared and annealed under the same conditions in order to make comparison of their properties reliable. It should be noted that data available in the literature suffer from significant scatter due to the strong dependence of the manganites properties to the synthetic route used and its thermal history.

# Experimental Section

All the samples were synthesized by solid state reaction starting from suitable amounts of $La_2O_3$ (Aldrich, 99.999%), $Mn_2O_3$ (Aldrich, 99.999%), $Na_2CO_3$ (Aldrich, 99.99%) and $CaCO_3$ (Aldrich, 99.99%). Pellets were prepared from well mixed powders and sintered at 1223 K for at least four days, during which time they were re-ground and re-pelletized at least twice. More details about sample preparation can be found in References 11 and 12.

X-ray powder diffraction patterns (XRPD) were acquired on a Bruker D8 Advance diffractometer equipped with a Cu anticathode. Electron microprobe analysis (EMPA) measurements were carried out using an ARL SEMQ scanning electron microscope equipped with energy dispersive detector, performing at least 10 measurements in different regions of each sample. Samples fragments were embedded in resin and their surface polished. Relative variance derived from these measurements is around 3%. According to EMPA and XRPD data, the above synthetic procedure gave single-phase materials; in addition each sample was found to be highly homogeneous in terms of chemical composition, with compositions close to the nominal ones.

In accordance with previous thermogravimetric measurements [9] the as prepared polycrystalline samples were annealed at 1123 K and $p(O_2) = 1$ atm to achieve the desired oxygen contents: $\delta = 0.12$ for $LaMnO_3$ and 0.08 for the other two samples. Mn-valence state can be inturn determined by considering the exact $\delta$–value and chemical composition of the samples.

Static magnetization was measured at 100 Oe from 350 K down to 2 K with a SQUID magnetometer (Quantum Design).

Electron paramagnetic resonance (EPR) measurements were performed at ~ 9.5 GHz with a Bruker spectrometer, with a continuous nitrogen flow used to study the temperature dependence in the range 160-470 K.

Neutron diffraction data were collected at the D1A instrument at the ILL Facility in Grenoble with a wavelength of 1.39 Å at RT. Neutron data were refined by means of FULLPROFILE software [29]. Parameters refined were: zero shift, scale factor, lattice constants, atomic parameters, fractional occupancies and isotropic thermal factors. The fractional occupancies for doped samples were refined by keeping a common parameter for both atoms, and with two separate parameters. The final results were in both cases very close one to each other.

# Results

Table 1 reports the refinement results for the RT neutron patterns of the three samples, while Figure 1a, b and c display the experimental pattern for the undoped, Na-doped and Ca-doped samples, respectively, together with the calculated pattern and the difference between them. Bragg peaks appear as vertical lines.

Neutron diffraction data reveal that a rhombohedral crystal structure (space group *R-3c*) is found for the pure and the Na-doped samples while an orthorhombic cell (S.G. *Pbnm*) is obtained for the Ca-doped sample; their lattice parameters are reported in Table 1. The cell volume, referred to the single unit cell (since Z is different for the two crystal structures), decreases in the order Na-doped sample (59.111 Å$^3$), pure sample (58.843 Å$^3$) and Ca-doped sample (58.693 Å$^3$). These small differences are consistent with the differences in the ionic radii between the ions; namely, 1.53 Å (Na), 1.50 Å (La) and 1.48 Å (Ca), thus confirming that for all the samples the Mn average oxidation state should be the same. We have to recall here that any variation of the Mn(III)/Mn(IV) ratio would lead to a major volume variation [12-14]. Structural data calculated from the refinements, namely Mn-O bond lengths and angles, are also reported in Table 1.

From the results of the refinements of the occupancies it is found that for the pure (undoped) sample the overall oxygen over-stoichiometry is compensated by more $V_{La}'''$ than $V_{Mn}'''$. Based on the neutron refinement results the correct formula for this sample is: La$_{0.952}$Mn$_{0.973}$O$_3$; this means that about 64% of the total cation vacancy population created by the oxygen over-stoichiometry are found on the lanthanum site.

For the Na-doped sample the actual formula, as derived from the neutron diffraction experiment, is (La$_{0.95}$Na$_{0.05}$)$_{0.976}$Mn$_{0.969}$O$_3$. In this case, the oxygen over-stoichiometry gives rise to a slightly lower occupancy of the B-site with a greater amount of $V_{Mn}'''$ (about 56% of the total number of vacancies).

Finally, for the Ca-doped sample, the refinement indicates that a greater population of cation vacancies (about 70%) is located on the La site; giving $(La_{0.9}Ca_{0.1})_{0.966}Mn_{0.986}O_3$.

Figure 2 reports the static molar magnetization collected at 100 Oe from 350 to 2 K for the three samples considered in this work. All the samples show an enhancement of the magnetization with decreasing temperature, as expected if a progressive evolution of the system from a paramagnetic (P) to a ferromagnetic (F) state occurs. The temperature of the P-F transition ($T_C$) was taken at the maximum of the first derivative of the molar magnetization ($M_{mol}$) *vs. T*, shown in the inset of Figure 2. Values of 146.8 K, 161.8 K and 141.8 K were obtained for pure, Ca-doped and Na-doped samples, respectively. By examining Figure 2 it is also clear that the P-F transition is sharper for the undoped and Ca-doped samples than for the Na-doped one. An estimate of the width of the magnetic transition can be made by determining the full width at half maximum of the magnetization derivative. In our case, the sharper transition corresponds to the Ca-doped sample (19 K) while a slight broadening can be detected for the undoped sample (20 K) and a more pronounced broadening for the Na-doped compound (35 K).

The value of the magnetization at the lowest investigated temperature decreases in the order Ca-doped sample, pure sample and Na-doped sample.

Finally, EPR investigation of all the samples in the range 160-470 K reproduced the main features already observed and discussed in detail in references 12 and 28 for Na and Ca doped manganites. Figure 3a shows the EPR spectra collected at RT for our samples. A unique signal with *g*-value $\cong 2$ is detected for the pure and Ca-doped samples, indicating the whole sample is paramagnetic at this temperature. For the Na doped sample a second component is present at lower resonant fields, in addition to the signal centred at $g \cong 2$, suggesting the presence of local inhomogeneous internal magnetic fields in some parts of the sample already at RT.

The temperature dependence of the EPR signals allows the magnetic behaviour to deepen, in particular when approaching the magnetic transition. Figure 3b reports the temperature dependence of the EPR spectrum of the Na-doped sample, as a selected example. At least two distinct signals are clearly detectable down to the minimum investigated temperature (160 K), while only for $T \geq 305$ K is a single signal with $g \cong 2$ present. This indicates the presence of different magnetic regions, in which the long-range magnetic interactions hold until different temperatures. A higher degree of magnetic homogeneity is instead shown by the temperature dependence of the EPR signal of the other two samples: for the Ca-doped sample a very weak second component is observed only for $T < 260$ K; for the pure sample a unique signal centred at $g \cong 2$ is observed over the whole investigated temperature range.

## Discussion

This work aims to show that among the various and important parameters affecting the physical properties of the manganites (La,M)MnO$_{3+\delta}$, a further one must be included, namely the cation vacancy distribution in the crystal lattice.

Let us consider the most relevant variables influencing the properties of the manganites. One of the fundamental variable, as commonly accepted, is the tolerance factor, $t$, defined as:

$$t = \frac{(r_A + r_O)}{\sqrt{2}(r_B + r_O)} \tag{2}$$

where $r_A$, $r_B$ and $r_O$ are the ionic radii (at RT and atmospheric pressure) for the corresponding ions in their appropriate coordination environments. The tolerance factor is indicative of the structural deformation in the perovskite structure arising from the misfit between different bond lengths. For the three samples considered the values of $t$ are very close to each other; 0.9776, 0.9781 and 0.9768 for the undoped, Na-doped and Ca-doped, respectively. The sample which deviates most from the ideal ($t = 1$) value for the cubic perovskite has the most distorted crystal structure of the compounds considered here, *i.e.* the Ca-doped system, which is orthorhombic. For the other two samples the structure is less distorted. The tolerance factor has previously been closely linked to the physical properties of mixed-valence manganites, namely transition temperatures in the magnetization and resistivity curves, since the level of structural distortion strongly affects the electronic behaviours of these materials. This was pointed out by Goodenough [31], who showed that the strength of ferromagnetic interaction between Mn(III) and Mn(IV) ions (competing with the Mn(t$^3$)-O(2p$_\pi$)-Mn(t$^3$) antiferromagnetic superexchange interaction) increases with the width $W_\sigma$ of the narrow σ* band formed by the O-2p$_\sigma$ and Mn-e$_g$ orbitals, which is given by:

$$W_\sigma \approx \varepsilon_\sigma \lambda_\sigma^2 \cos\phi \cos\theta_{ij} \tag{3}$$

where $\varepsilon_\sigma$ represents the stabilization energy of the band, $\lambda_\sigma$ is the overlap integral between atomic orbitals and constitutes a measure of the mixing between the Mn-$e_g$ and the O-$2p_\sigma$ orbitals; $\phi$ is the deviation of the Mn-O-Mn bond angle with respect to 180° [180-$\phi$], while $\theta_{ij}$ is the angle between the spins on neighboring Mn atoms. Oxidation of the MnO$_3$ array increases both $\lambda_\sigma$ and cos$\phi$, ferromagnetic ordering below $T_C$ increases cos$\theta_{ij}$. In all the samples considered here the mean oxidation state of Mn ions is the same, which allows us to focus attention on the role of cation vacancies and the structural data.

From the discussion just presented, it is clear that an orthorhombic deformation of the unit cell should give rise to weakened magnetic properties with respect to a less distorted structure. This is mainly due the fact that: i) the Mn-O-Mn bond angle(s) is less close to the "ideal" value of 180°, and ii) the Mn-O bond lengths are in general greater, and have three different values, compared to the single value of the rhombohedral structure. The mean value of the Mn-O bond length is 1.970(1) Å for the Ca-doped sample and decreases to 1.969(1) Å in the Na-doped sample and 1.967(1) Å in the undoped system. In addition, the angle for the Ca-doped sample (considering the in-plane angle) is the lowest between the three samples (161.62°) while the highest is found in the Na-doped sample (163.18°).

Therefore, based on all these considerations we should expect the lowest $T_C$ to be exhibited by the Ca-doped sample; instead, it displays the highest $T_C$ of the three samples (161.8 K). It is clear that the usual parameters employed to describe the trends in the physical properties of manganites fail here.

Vergara *et al.* [13] first recognized that in order to characterize completely the mixed-valence manganites a "new" parameter was needed: *the concentration of vacancies on the perovskite B site* [13]. In that study, however, the authors simply calculated the vacancy concentration on the B site from the oxygen stoichiometry assuming an equal distribution over A and B sites in the structure. This can of course give rise to a clear trend if a single series of compounds is analyzed, as found in the few studies which considered the role of cation vacancy concentration. However, we will show that in order to correctly characterize the differently doped manganites in a complete way a further parameter is needed; viz. *the cation vacancy distribution on the A and B sites*.

Let us consider the three samples studied here. If we assume the percentage of vacancies on the B-site as the total cation vacancy concentration derived from neutron diffraction measurements divided by two, we will obtain what is usually considered to be the "vacancy concentration on the B site" which in the following discussion will be considered as the *formal* value of $V'''_{Mn}$. In Figure 4 we have plotted the Curie temperatures against this quantity for the three samples (full symbols). A trend similar to those reported in the literature [13,32] for a series of samples with various *x* values but with the same dopant ion, *i.e.* a progressive reduction of $T_C$ as the concentration of $V'''_{Mn}$ increases, is not found from our data if we plot them against the *formal* value of $V'''_{Mn}$. In order to find a direct correlation with the vacancy concentration for samples with different dopants we plotted the $T_C$ values against the *real* vacancy concentration on the B-site as *directly* determined from the refinement of the neutron data (open symbols, Figure 4). As can be seen from the plot, the data now fall on a straight line.

In addition to the value of the transition temperatures the character of the magnetic transition also seems to be directly dependent on the actual concentration of vacancies on the B site. This can be concluded by examining the curves in Figure 2 and at the width of the transitions; it is clear that by increasing $V'''_{Mn}$ both the $T_C$ decreases and the transitions become wider. We also note that the value of the molar magnetization measured at 2 K decreases as the $V'''_{Mn}$ concentration increases. These results can be directly correlated to the fact that as the $V'''_{Mn}$ concentration increases the system tends to contain more clusters separated by regions where the cation vacancies break the magnetic interactions between neighbouring Mn ions and where localization of carriers occurs.

The EPR results also agree with this interpretation. As explained in detail by Malavasi *et al.* [12], the multiplicity of the EPR signals indicates the temperature region in which the onset and completion of the magnetic transition occurs and the extent of this temperature region corresponds to the broadening of the magnetic transition itself. The multiplicity of the EPR signal observed for the Na-doped sample over a wide temperature range suggests the presence of regions with different long-range magnetic interactions, arising from a relatively high Mn vacancy concentration. This is not the case for the other two samples, in agreement with the sharp P-F transition and for which a lower $V_{Mn}'''$ concentration is found.

Finally, taking into account that the EPR signal has the same origin [30] for each of these three systems, we plotted the EPR signal intensity (area) values against the *formal* value of $V_{Mn}'''$ and against the *real* value of $V_{Mn}'''$ (see the inset of Figure 4), as already analyzed for the $T_C$ values (Figure 4). For this purpose we considered the spectra collected at the highest investigated temperature (470 K), where all the samples are paramagnetic, taking into account the dependence of the paramagnetic susceptibility on $1/(T-T_C)$. It is clear that the EPR intensities progressively decrease with increasing the Mn vacancies concentration only when considering the real vacancy concentration, $V_{Mn}'''$.

A clear conclusion from this work is therefore that in order to correctly compare and evaluate the physical properties of $LaMnO_{3+\delta}$ manganites we need to consider not only the cation vacancy concentration, but also their relative distribution between the A and B sites of the perovskite structure.

Let us consider now how vacancies are distributed in the lattice for the different samples. If we look at the relative percentage of $V_{Mn}'''$ with respect to the total vacancies it is seen that this value increases from 30% for the Ca-doped sample to 36% for the undoped system and to 56% for the Na-doped sample. Although a simple correlation between these data cannot be found, we attempt to provide some indication of the possible features that can influence their values. We recall here that previous atomistic simulation studies suggest that in both orthorhombic and rhombohedral structures oxidative non-stoichiometry leads to the formation of cation vacancies on both La and Mn sites, with a tendency towards more La vacancies [24]. From our results, we note that for the rhombohedral samples the amount of $V_{Mn}'''$ is greater than in the orthorhombic sample. In addition, as the difference in the $V_{Mn}'''$ distribution between the two rhombohedral manganites is not a structural effect, it can be probably linked to the type of cation present on the A site. In the following discussion we try to suggest a possible origin for this difference. Figure 5 reports the concentration of vacancies on the B site as a function of the mean ionic radius on the A site ($<r_A>$) of the perovskite structure. This plot suggests that a degree of correlation exists; as the average ionic radius on the A site increases, the amount of vacancies on the B site also increases. This can in turn be linked to the reduced tendency of vacancies to form on the A site as a consequence of the higher energy required for its formation when the mean ionic radius increases due to the reduction of the defect polarization energy, which is given by [33]:

$$u_p = \frac{e^2}{2r}\left(1 - \frac{1}{\varepsilon}\right) \qquad (4)$$

where $r$ is the (average) ionic radius, $<r_A>$, of the cation and $\varepsilon$ the dielectric constant of the medium. If we consider that the overall nature of the ions on the B site is not greatly affected by the substitution because the Mn ions have the same average oxidation state, we can expect a stronger effect on the A site. Other effects probably play a role in determining the relative energetics of defect formation. However, these results suggest that in order to reduce the tendency of Mn-vacancies to form, smaller ions could be preferred on the A site.

In addition, we would like to stress that the correlation between the Curie temperatures and $<r_A>$ found in this work is the opposite to that usually reported in the literature, as can be seen in Figure 5 (open symbols). The trend usually reported for $T_C$ is based on the consideration that, within the same series of samples, the increase of the ionic radius on the A site contributes to an increase in the width of the conduction band and consequently of the magnetic coupling. However, we have already demonstrated that the geometrical (structural) data alone are not able to account for the trend observed when dealing with oxygen over-stoichiometric samples with different dopants.

In any case, in order to obtain other evidence for our findings we plan to investigate this system using computer simulation techniques and to enrich the experimental data by working with Sr- and K-doped samples since the actual number of samples considered in this work is sufficient to rise the question about the role of vacancy distribution in manganites but maybe not high enough to extract a fully conclusive picture of this problem.

# Conclusion

In the present paper we have correlated the structural information gained by neutron diffraction measurements, particularly the cation vacancy distribution over A and B sites, to the magnetic properties of doped manganites, $(La,M)MnO_{3+\delta}$, where M = Na, Ca. This work shows that, in order to fully characterize the properties of these manganites and give a meaningful interpretation to physical quantities, a more complete knowledge of the crystal stoichiometry and defect chemistry is required. Moreover, due to the strong dependence of the Curie temperatures and other magnetic properties on the preparation, thermal history and annealing processes (both equilibrium and non-equilibrium), we can also expect a strong dependence of cation vacancy distribution from these synthesis treatments. It therefore seems that systematic investigations of the cation vacancy distribution over A and B sites of perovskite manganites as a function of several physicochemical parameters is crucial for a better understanding of their properties, and should assist in rationalizing some previously published data. Finally, we would like to stress that this study may not be fully conclusive but we have attempted to highlight a key question in CMR manganites.

# Acknowledgement

Financial support from the Italian Ministry of Scientific Research (MIUR) by PRIN Projects (2004) is gratefully acknowledged.

# References

1. R. Mahendiran, S.K. Tiwary, A.K. Raychaudhuri, T.V. Ramakrishnan, R. Mahesh, N. Rangavittal and C.N.R. Rao, *Phys. Rev. B* 53 (1996) 3348.

2. P. Schiffer, A.P. Ramirez, W. Bao and S.W. Cheong, *Phys. Rev. Lett.* 75 (1995) 3336.

3. C. Zener, *Phys. Rev.* 82 (1951) 403.

4. A.J. Millis, P.B. Littelwood and B.I. Shraiman, *Phys. Rev. Lett.* 74 (1995) 5144.

5. H. Röder, Jun Zang and A.R. Bishop, *Phys. Rev. Lett.* 76 (1996) 1356.

6. K.H. Ahn, T. Lookman, A.R. Bishop, *Nature* 428 (2004) 401.

7. A. Prodi, E. Gilioli, A. Gauzzi, F. Licci, M. Marezio, F. Bolzoni, Q. Huang, A. Santori, and J.W. Lynn, *Nature Mater.* 3 (2004) 48.

8. C.N.R. Rao, *J. Phys. Chem. B.* 104 (2000) 5877.

9. L.M. Rodriguez, and J.P. Attfield, *Chem. Mater.* 11 (1999) 1504.

10. Y. Murakami, J.H. Yoo, D. Shindo, T. Aton, and M. Kikuchi, *Nature* 423 (2003) 965.

11. L. Malavasi, M.C. Mozzati, C.B. Azzoni, G. Chiodelli, G. Flor, *Solid State Commun.* 123 (2002) 321.

12. L. Malavasi, M.C. Mozzati, P. Ghigna, C.B. Azzoni, G. Flor, *J. Phys. Chem. B* 107 (2003) 2500.

13. J. Vergara, R.J. Ortega-Hertogs, V. Madruga, F. Sapiña, Z. El-Fadii, E. Martinez, A. Beltrán and K.V. Rao, *Phys. Rev. B* 60 (1999) 1127.

14. J. Töpfer, and J.B. Goodenough, *Chem. Mater.* 9 (1997) 1467.

15. E.T. Maguire, A.M. Coats, J.M.S. Skakle, and A.R. West, *J. Mater. Chem.* 9 (1999) 1337.

16. G. Dezanneau, A. Sin, H. Roussel, M. Audier, and H. Vincent, *J. Solid State Chem.* 173 (2003) 216.

17. K. Nakamura, *J. Solid State Chem.* 173 (2003) 299.

18. B. Raveau, A. Maignan, C. Martin and M. Hervieu, *Chem. Mater.* 10 (1998) 2641.

[19] E. Herrero, J. Alonso, J.L. Martinez, M. Vallet-Regt and J.M. Gonzales-Calbet, *Chem. Mater.* 12 (2000) 1060.

[20] C. Ritter, M.R. Ibarra, J.M. De Teresa, P.A. Algarabel, C. Marquina, J. Blasco, J. Garcia, S. Oseroff and S-W. Cheong, *Phys. Rev. B* 56 (1997) 8902.

[21] A. Arulraj, R. Mahesh, G.N. Subbanna, R. Mahendiran, A.K. Raychaudhuri, C.N.R. Rao, *J. Solid State Chem.* 127 (1996) 87.

[22] J.A. Alonso, M.J. Martinez-Lope, M.T. Casais, J.L. MacManus-Driscoll, P.S.I.P.N. de Silva, L.F. Cohen and M.T. Fernandez-Diaz, *J. Mater. Chem.* 7 (1997) 2139.

[23] J.A.M. van Roosmalen and E.H.P. Cordfunke, *J. Solid State Chem.* 93 (1991) 212.

[24] J.A.M. van Roosmalen, E.H.P. Cordfunke, R.B. Helmholdt and H.W. Zandbergen, *J. Solid State Chem.* 110 (1994) 100.

[25] R.A. De Souza, M.S. Islam and E. Ivers-Tiffée, *J. Mater. Chem.* 9 (1999) 1621.

[26] M. Hervieu, European *J. Solid State Inor. Chem.* 32 (1995) 79.

[27] J.F. Mitchell, D.N. Argyriou, C.D. Potter, D.G. Hinks, J.D. Jorgensen and S.D. Bader, *Phys. Rev. B* 54 (1996) 6172.

[28] Q. Huang, A. Santoro, J.W. Lynn, R.W. Erwit, J.A. Borchers, J.L. Peng and R.L. Greene, *Phys. Rev. B* 55 (1997) 14987.

[29] J. Rodriguez-Carvajal, Physica B 192 (1993) 55.

[30] S.B. Oseroff, M. Torikachvili, J. Singley, S. Ali, S-W. Cheong and S. Schultz, *Phys. Rev. B* 53 (1996) 6521.

[31] J.B. Goodenough, Localized to itinerant electronic transition in perovskite oxides, *Structure and Bonding*, Springer-Verlag Berlin, 2001, 98.

[32] T. Boix, F. Sapina, Z. El-Fadli, E. Martinez, A. Beltran, J. Vergara, R.J. Ortega, K.V. Rao, *Chem. Mater.* 10 (1998) 1569.

[33] H. Schmalzried, *Solid State Reactions*, Verlag Chemie, Basel 1981, 25.

**Table 1**

|  | Structural Parameters | $La_{0.961}Mn_{0.961}O_3$ | $(La_{0.95}Na_{0.05})_{0.974}Mn_{0.974}O_3$ | $(La_{0.9}Ca_{0.1})_{0.974}Mn_{0.974}O_3$ |
|---|---|---|---|---|
|  | $a$/Å | 5.5287(1) | 5.5331(2) | 5.5166(5) |
|  | $b$/Å | 5.5287(1) | 5.5331(2) | 5.4833(5) |
|  | $c$/Å | 13.3368(1) | 13.3768(3) | 7.7598(7) |
|  | $V$/ Å$^3$ | 58.843(3) | 59.111(7) | 58.693(2) |
| A-site | Occ. | 0.952(2) | 0.976(2) | 0.966(1) |
|  | $x$ | - | - | -0.0037(2) |
|  | $y$ | - | - | 0.0183(3) |
| B - site | Occ. | 0.973(3) | 0.969(2) | 0.986(4) |
| O(1) | $x$ | 0.4469(1) | 0.4479(1) | 0.0635(2) |
|  | $y$ | - | - | 0.4935(2) |
| O(2) | $x$ | - | - | 0.7278(2) |
|  | $y$ | - | - | 0.2733(3) |
|  | $z$ | - | - | 0.0335(2) |
|  | $R_{wp}$/GoF | 5.38/1.40 | 4.90/1.37 | 5.19/1.32 |
|  | Mn-O/Å | 1.967(1) | 1.969(1) | 1.973(1) (*eq*) |
|  |  |  |  | 1.966(1) (*eq*) |
|  |  |  |  | 1.972(2) (*ax*) |
|  | Mn-O-Mn/° | 162.85 | 163.18 | 161.62 O(2) |
|  |  |  |  | 159.424 O(1) |
| A-site | $B_{iso}$ | 0.83(2) | 0.79(2) | 0.88(2) |
| B-site | $B_{iso}$ | 0.56(3) | 0.71(3) | 0.65(3) |
| O(1) | $B_{iso}$ | 1.38(2) | 1.28(2) | 1.06(4) |
| O(2) | $B_{iso}$ | - | - | 1.30(3) |

# Figures Captions

**Figure 1** – Neutron diffraction pattern for the undoped (A), Na-doped (B) and Ca-doped (C) samples. Red crosses represent the experimental pattern, the black line is the calculated one, while the blue line is the difference between them. Bragg peaks appear as vertical green lines. The inset shows an enlargement of a small portion of the pattern. Pattern regions removed from the refinement included sample holder contribution.

**Figure 2** – Molar magnetization at 100 Oe ($M_{mol}$) *vs.* temperature ($T$) for the three samples: undoped (grey squares), Na-doped (open squares) and Ca-doped (full squares). In the inset: first derivative of $M_{mol}$ *vs.* $T$ for the three samples (same symbols as before).

**Figure 3** – (a) EPR spectra of the three samples collected at 295 K. (b) Temperature dependence of EPR signal for the Na-doped sample.

**Figure 4** – Curie temperatures ($T_C$) *vs.* the *formal* $V_{Mn}'''$ concentration (full symbols) and *vs.* the *real* concentration (open symbols). Circles, triangles-down and squares represent the undoped, Na-doped and Ca-doped samples, respectively. In the inset the EPR intensity at 470 K is plotted *vs.* the *formal* $V_{Mn}'''$ and *vs.* the *real* $V_{Mn}'''$ concentration (same symbols as before).

**Figure 5** – Relative percentage of $V_{Mn}'''$ concentration (full symbols) and Curie temperatures (open symbols) *vs.* the mean ionic radius of the A site cation, $<r_A>$. Circles, triangles-down and squares represent the undoped, Na-doped and Ca-doped samples, respectively.

# Table Caption

**Table 1 –** Rietveld refinement results for the three samples considered in this work. The formula are written considering an equal vacancy concentration between the A and B sites.

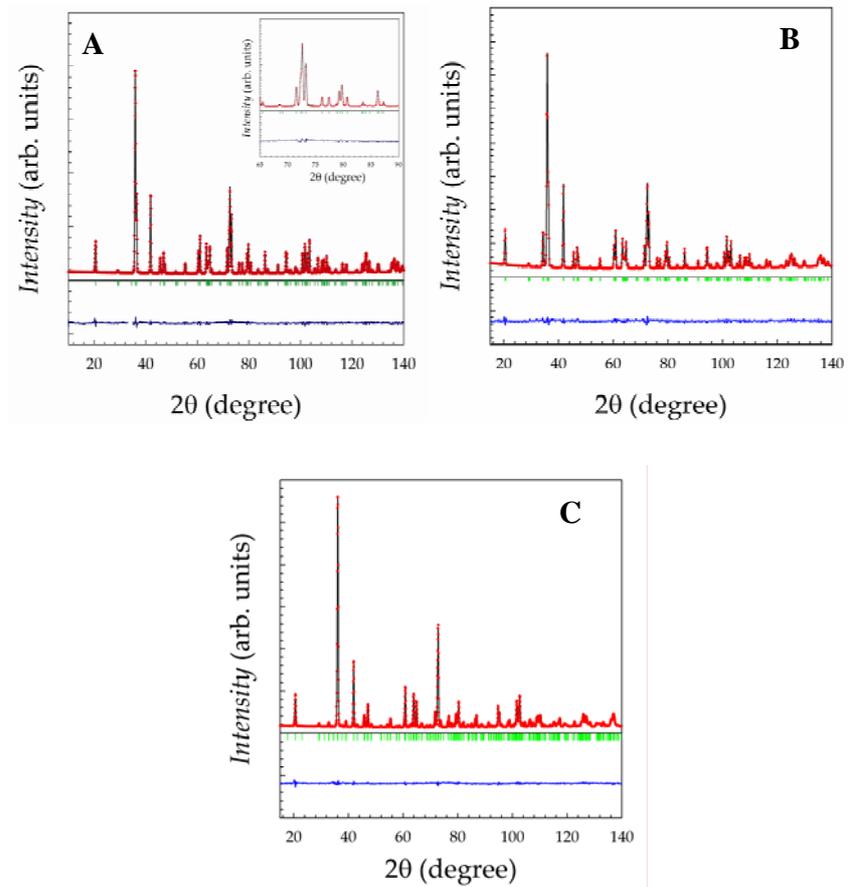

Figure 1

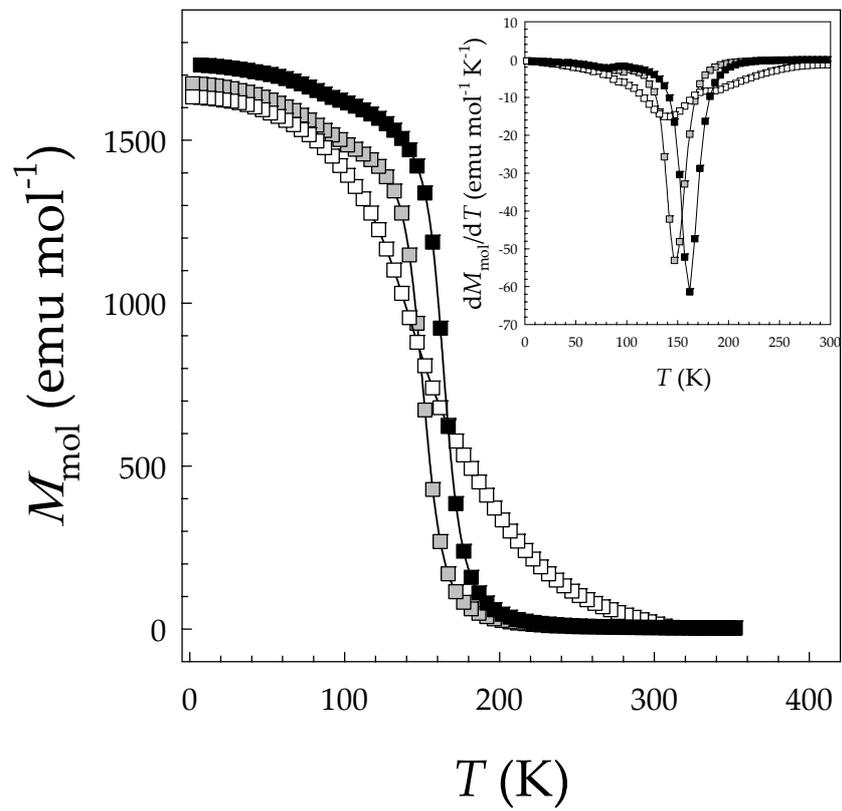

Figure 2

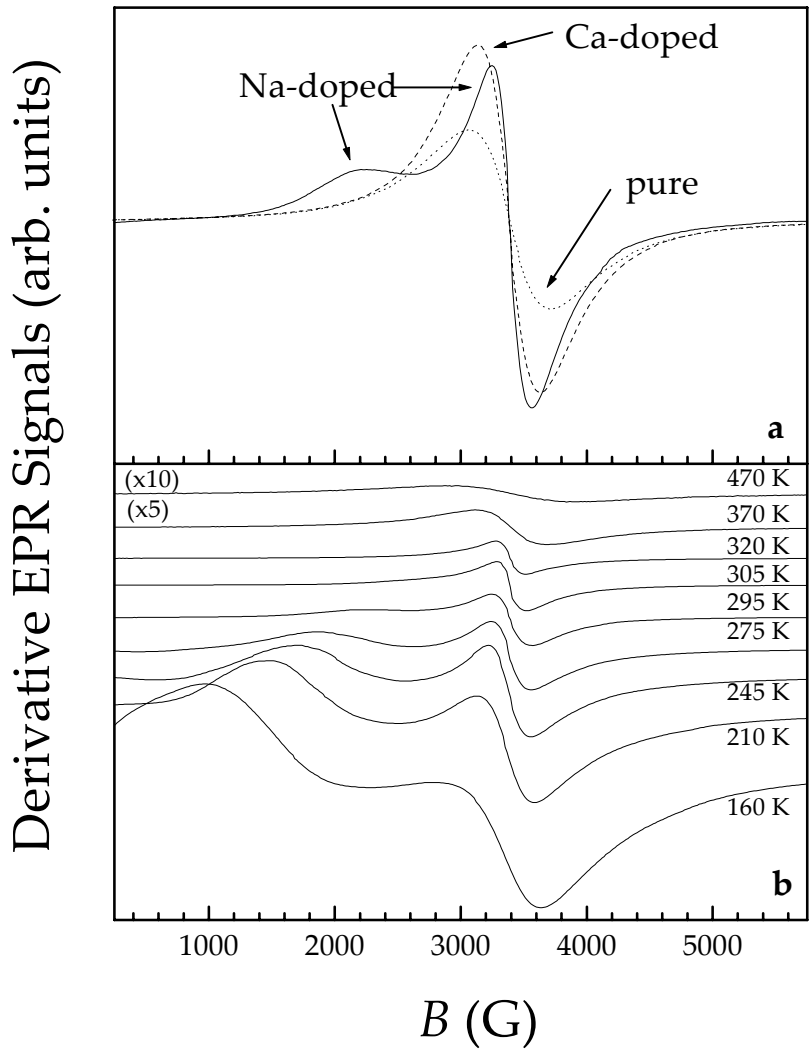

Figure 3

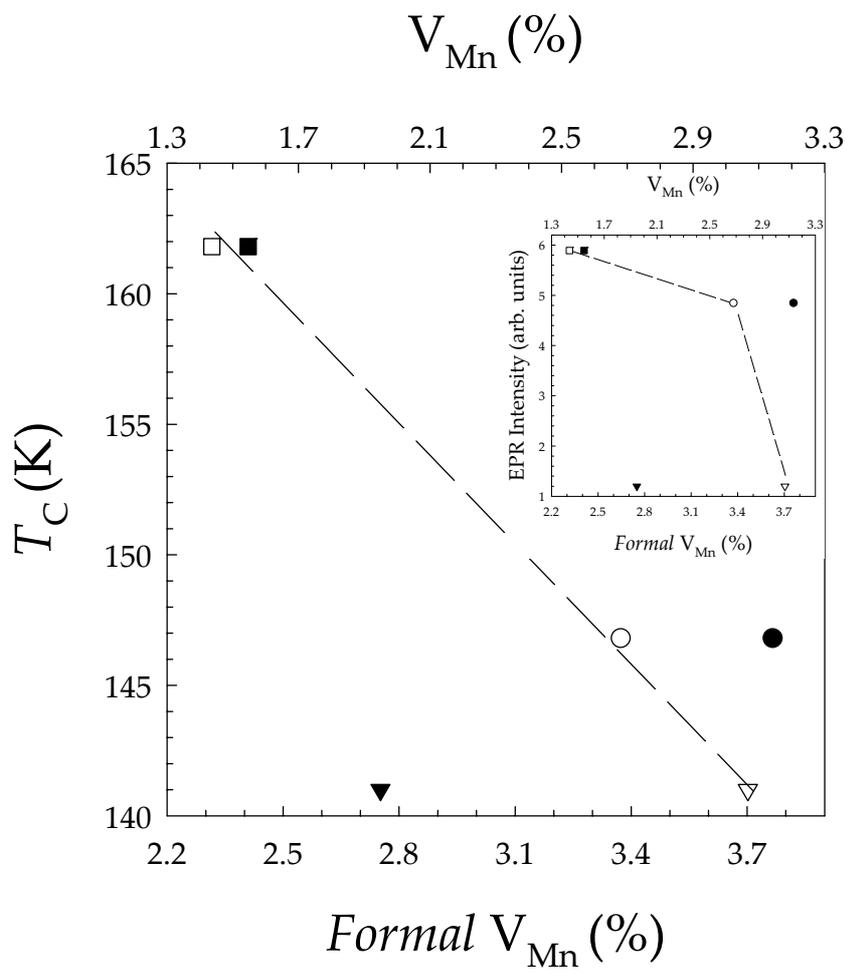

Figure 4

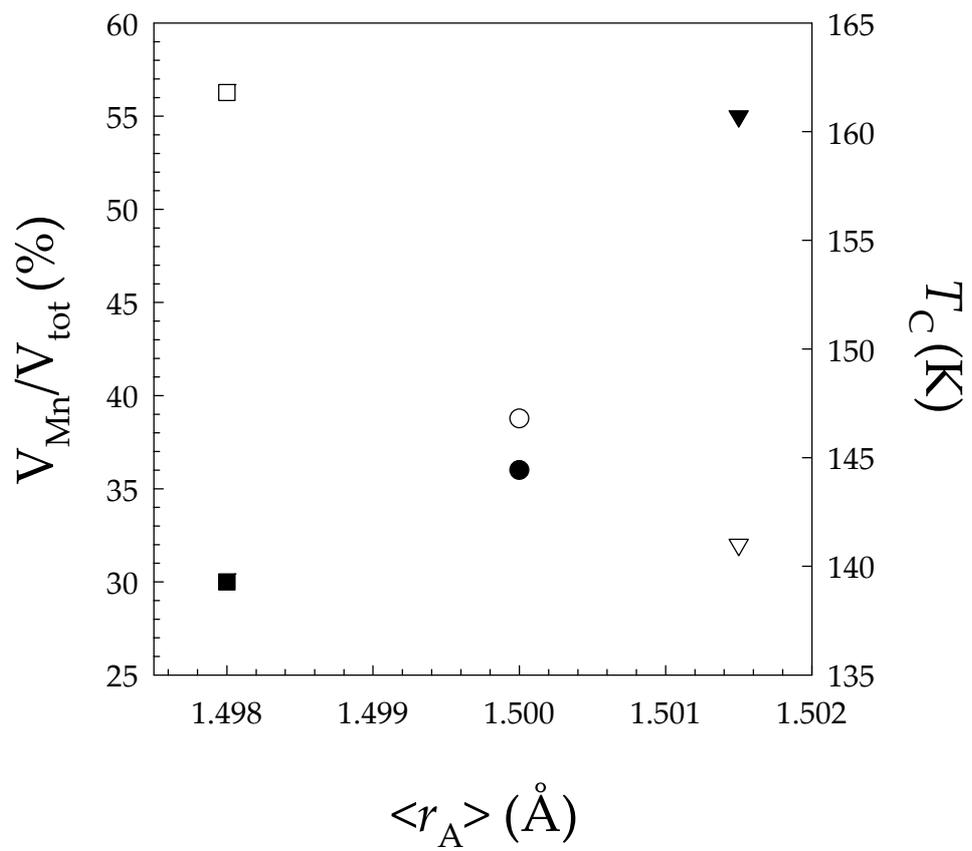

Figure 5